\documentclass[11pt]{article}
\usepackage{geometry}
\usepackage{color}
\usepackage{amsmath}
\usepackage{amssymb}
\usepackage{graphicx}

\makeatletter
\newcommand{\lyxaddress}[1]{
	\par {\raggedright #1
	\vspace{1.4em}
	\noindent\par}
}

\@ifundefined{date}{}{\date{}}
\usepackage{amsfonts}
\usepackage{graphicx}
\providecommand{\U}[1]{\protect\rule{.1in}{.1in}}

\makeatother

\begin{document}

\title{\textbf{Inflationary dynamics in modified gravity models}}

\author{R. R. Cuzinatto$^{1}$ and L. G. Medeiros$^{2}$}
\maketitle

\lyxaddress{$^{1}$Instituto de Ciência e Tecnologia, Universidade Federal de
Alfenas, Rodovia José Aurélio Vilela, 11999, CEP 37715-400, Poços
de Caldas, MG, Brazil}

\lyxaddress{$^{2}$Escola de Ciência e Tecnologia, Universidade Federal do Rio
Grande do Norte, Campus Universitário, s/n--Lagoa Nova, CEP 59078-970,
Natal, RN, Brazil}
\begin{abstract}
Higher-order theories of gravity are a branch of modified gravity
wherein the geometrodynamics of the four-dimensional Riemannian manifold
is determined by field equations involving derivatives of the metric
tensor of order higher than two. This paper considers a general action
built with the Einstein-Hilbert term plus additional curvature-based
invariants, viz. the Starobinsky $R^{2}$-type term, a term scaling
with $R^{3}$, and a correction of the type $R\square R$. The focus
is on the background inflationary regime accommodated by these three
models. For that, the higher-order field equations are built and specified
for the FLRW line element. The dynanical analysis in the phase space
is carried in each case. This analysis shows that the Starobinsky-plus-$R^{3}$
model keeps the good features exhibited by the pure Starobinsky inflationary
model, although the set of initial conditions for the inflaton field
$\chi$ leading to a graceful exit scenario is more contrived; the
coupling constant $\alpha_{0}$ of the $R^{3}$ invariant is also
constrained by the dynamical analysis. The Starobinsky-plus-$R\square R$
model turns out being a double-field inflation model; it consistently
enables an almost-exponential primordial acceleration followed by
a radiation dominated universe if its coupling $\beta_{0}$ takes
values in the interval $0\leq\beta_{0}\leq3/4$. The models introducing
higher-order correction to Starobinsky inflation are interesting due
to the possibility of a running spectral index $n_{s}$, something
that is allowed by current CMB observations.
\end{abstract}

\section{Introduction \label{sec:Introduction}}

General relativity (GR) currently stands as the canonical theory describing
the gravitational interaction. Since its proposition early in the
XX century, GR was able to explain and predict a plethora of phenomena
in the realms of physics, astrophysics and cosmology. Among them are
the examples of gravitational redshift \cite{Weinberg1972}, gravitational
lensing \cite{Bartelmann2010}, prediction of existence of black holes
\cite{Wald1984} and gravitational waves \cite{Maggiore2008}, and
the description of the universe's large scale evolution \cite{Padmanabham1993}.
Even so, there are indications that GR is not a definitive theory
of gravity. The hints are structural in nature---e.g. the existence
of singularities within GR---or particularly related to high-energy
regimes: GR can not be trivially quantized \cite{Utiyama1962,Wood2009}
and it does not provide a completely consistent description of the
primeval universe (around the energy scales related to inflation)
\cite{Guth1981,Starobinsky1980}. Therefore, it is only natural to
propose modification to GR in an attempt to overcome these challenges.

From a purely theoretical point of view, GR is built by considering
that gravity is described by a metric-compatible four-dimensional
Riemannian manifold, which is endowed with a single rank-2 tensorial
field---the metric tensor $g_{\mu\nu}$---, which is invariant under
diffeomorphisms, and which exhibits second-order equations of motion
(cf. the Lovelock theorem) \cite{Clifton2012}. Modifications to GR
are implemented by relaxing anyone of the aforementioned hypotheses.
For instance, Horndeski theories \cite{Horndesk1974} stem from violating
the hypothesis that the metric is the only fundamental field: an extra
degree of freedom is also assumed. A different pathway is to admit
a Riemann-Cartan-type of spacetime substrate, a manifold equipped
with an affine connection bearing a non-null antisymmetric sector;
in this case, torsion is included as a gravitational entity and the
Einstein-Cartan theories are born \cite{Trautman2006}. Another possibility
is to eliminate curvature while keeping a non-null torsion; this is
a feature of Weitzenböck manifold and the teleparallel equivalent
of general relativity \cite{Aldrovandi2013,Pereira2019,Krssak2019,Aldrovandi2004,Aldrovandi2003,Andrade2000}.\footnote{Regarding the teleparallel framework for gravity, we also point the
reader to the contribution by P. J. Pompeia for this book and the
references cited in that paper.} On the other hand, if the fields equations for the metric tensor
are allowed to include derivatives of order greater than two---while
simultaneously maintaining all the other hypotheses---then the higher-order
gravity theories are obtained \cite{EPJC2008,PRD2016}.

Higher-order theories of gravity feature additional terms to the Einstein-Hilbert
(EH) action engendering higher-order derivatives in the field equations.
Such extra terms in the action may be seen as correction terms, classified
according to their typical mass/energy scale. Following this classification,
zero-order terms are those counted in units of square mass; they correspond
to the curvature scalar $R$ and the cosmological constant $\Lambda$
in the EH action. First-order corrections to EH action involve term
of mass to the fourth power; these are built with the invariants
\begin{equation}
R^{2}\qquad\text{and}\qquad R_{\mu\nu}R^{\mu\nu}.\label{eq:1st-order}
\end{equation}
It is worth mentioning that the other two possible first-order invariants,
$R_{\mu\nu\alpha\beta}R^{\mu\nu\alpha\beta}$ and $\square R$, do
not contribute to the field equations.\footnote{The term $R_{\mu\nu\alpha\beta}R^{\mu\nu\alpha\beta}$ may be written
as a linear combination of $R^{2}$, $R_{\mu\nu}R^{\mu\nu}$, and
the Gauss-Bonnet topological invariant. The term $\square R$ is explicitly
a surface term.} Second-order terms are corrections to EH action having units of mass
to the sixth power; they made up with the following invariants \cite{Decanini2007}

\begin{gather}
R\square R,\,R_{\mu\nu}\square R^{\mu\nu},\,\nonumber \\
R^{3},\,RR_{\mu\nu}R^{\mu\nu},\,R_{\mu\nu}R_{\hphantom{\text{\ensuremath{\nu}}}\alpha}^{\nu}R^{\alpha\mu},\,\label{eq:2nd-order}\\
RR_{\mu\nu\alpha\beta}R^{\mu\nu\alpha\beta},\,R_{\mu\alpha}R_{\nu\beta}R^{\mu\nu\alpha\beta},\,\text{and}\,R_{\mu\nu\alpha\beta}R_{\hphantom{\alpha\beta}\kappa\rho}^{\alpha\beta}R^{\kappa\rho\mu\nu}.\nonumber
\end{gather}

Among the various applications of higher-order theories of gravity
\cite{GERG2015,PRD2020,PRD2021,PRD2023}, one class of particular
interest is that of inflationary cosmology \cite{PRD2019,JCAP 2019,PRD2022,EPJC2023}.

In the end of 1979, Alexei A. Starobinsky proposed that quantum gravitational
effects, presumably significant in the primordial universe, produce
a quasi-de Sitter cosmic dynamics, i.e. an almost-exponential inflationary
regime \cite{Starobinsky1980,Starobinsky1979}. In fact, A. A. Starobinsky
showed that the inclusion of the term $R^{2}$ in the EH action is
able to generate an early accelerated expansion ending in a radiation-dominated
decelerated universe. Starobinsky model is an enormous success: nowadays,
it is one of the most promising candidates for realizing the inflationary
dynamics. The main reason for this accomplishment is its being a single-parameter
model fitting perfectly the most recent observations of the cosmic
microwave background radiation (CMB) \cite{Planck2020,BICEP3}. Moreover,
the theoretical motivation for Starobinsky model is quite robust.
In effect, Starobinsky inflation occurs in energy scales of about
$10^{15}\text{ GeV}$; in such period the action containing the term
$R^{2}$ may be considered as part of a higher-order theory expected
in the context of quantization of gravity \cite{Stelle1977}.

The main goal of this contribution is to review the basic aspects
of the cosmic dynamics predicted by Starobinsky inflation and to study
its extension to models containing second-order derivative corrections
involving the curvature scalar. Section \ref{sec:Higher-order-model}
presents a general action integral encompassing the regular EH term,
plus Starobinsky $R^{2}$-contributions, and the novel higher-order
corrections; the field equations for this modified gravity are also
derived therein. Section \ref{sec:Inflation} summarizes the conditions
for inflation in a homogeneous and isotropic background; the field
equations are also specified in FLRW spacetime. Subsections \ref{subsec:Starobinsky},
\ref{subsec:Starobinsky-R3}, and \ref{subsec:Starobinsky-Rd2R} analyse
the inflationary dynamics (in the phase space) in three separate models,
viz. the original Starobinsky proposal, the model supplementing Starobinsky
term with a $R^{3}$ contribution, and a higher-order model adding
a correction of the type $R\square R$ to the traditional $R^{2}$-term.
Section \ref{sec:Conclusion} brings our final comments.

\section{ Fundamentals of the proposed modified gravity models \label{sec:Higher-order-model}}

The most general action presenting up to second order correction to
the EH action involving the curvature scalar reads:
\begin{equation}
S=\frac{M_{\text{Pl}}^{2}}{2}\int d^{4}x\sqrt{-g}\left[R+\frac{1}{2\kappa_{0}}R^{2}+\frac{\alpha_{0}}{3\kappa_{0}^{2}}R^{3}-\frac{\beta_{0}}{2\kappa_{0}^{2}}R\square R\right].\label{eq:S}
\end{equation}
Herein $\kappa_{0}$ has units of square mass while $\alpha_{0}$
and $\beta_{0}$ are dimensionless parameters. Starobinsky $R^{2}$
term introduce the first-order correction to Einstein-Hilbert $R$
term. The last two terms of (\ref{eq:S}) account for all the possible
second-order corrections built with the curvature scalar. Parameter
$\kappa_{0}$ sets the energy scale for inflation; $\alpha_{0}$ and
$\beta_{0}$ regulate the deviations from Starobinsky model.

It is convenient to perform a conformal metric transformation and
to introduce dimensionless fields as follows:

\begin{equation}
\bar{g}_{\mu\nu}=e^{\chi}g_{\mu\nu},\quad\lambda=\frac{R}{\kappa_{0}}\quad\text{and}\quad e^{\chi}=1+\lambda+\alpha_{0}\lambda^{2}-\frac{\beta_{0}}{\kappa_{0}}\square\lambda.\label{eq:field-redefinitions}
\end{equation}
The above allows one to cast (\ref{eq:S}) in the Einstein frame \cite{EPJC2023}:
\begin{equation}
\bar{S}=\frac{M_{\text{Pl}}^{2}}{2}\int d^{4}x\sqrt{-\bar{g}}\left[\bar{R}-3\left(\frac{1}{2}\bar{\nabla}_{\rho}\chi\bar{\nabla}^{\rho}\chi-\frac{\beta_{0}}{6}e^{-\chi}\bar{\nabla}_{\rho}\lambda\bar{\nabla}^{\rho}\lambda+V\left(\chi,\lambda\right)\right)\right],\label{eq:S-Einstein-frame}
\end{equation}
where
\begin{equation}
V\left(\chi,\lambda\right)=\frac{\kappa_{0}}{3}e^{-2\chi}\lambda\left(e^{\chi}-1-\frac{1}{2}\lambda-\frac{\alpha_{0}}{3}\lambda^{2}\right),\label{eq:potential}
\end{equation}
stands for the multi-field potential of our model. The latter is a
gravity model described in terms of the metric tensor $\bar{g}_{\mu\nu}$
along with two scalar fields, viz. $\chi$ and $\lambda$.

The field equations follow from setting to zero the variations of
the action (\ref{eq:S-Einstein-frame}) with respect to the fields
$\bar{g}_{\mu\nu}$, $\chi$ and $\lambda$. Executing this procedure
for the metric tensor yields:
\begin{equation}
\bar{R}_{\mu\nu}-\frac{1}{2}\bar{g}_{\mu\nu}\bar{R}=\frac{1}{M_{\text{Pl}}^{2}}\bar{T}_{\mu\nu}^{\left(\text{eff}\right)},\label{eq:Einstein-FE}
\end{equation}
with the effective energy momentum tensor given by
\begin{equation}
\frac{1}{M_{\text{Pl}}^{2}}\bar{T}_{\mu\nu}^{\left(\text{eff}\right)}=\frac{3}{2}\left(\bar{\nabla}_{\mu}\chi\bar{\nabla}_{\nu}\chi-\frac{1}{2}\bar{g}_{\mu\nu}\bar{\nabla}^{\rho}\chi\bar{\nabla}_{\rho}\chi\right)-\frac{\beta_{0}e^{-\chi}}{2}\left(\bar{\nabla}_{\mu}\lambda\bar{\nabla}_{\nu}\lambda-\frac{1}{2}\bar{g}_{\mu\nu}\bar{\nabla}^{\rho}\lambda\bar{\nabla}_{\rho}\lambda\right)-\frac{3}{2}\bar{g}_{\mu\nu}V\left(\chi,\lambda\right).\label{T_munu efetivo}
\end{equation}
The field equations for the scalar fields are:
\begin{gather}
\bar{\square}\chi-\frac{\beta_{0}}{6}e^{-\chi}\bar{\nabla}_{\rho}\lambda\bar{\nabla}^{\rho}\lambda-V_{\chi}=0,\label{eq:chi-FE}\\
\beta_{0}e^{-\chi}\left(\bar{\nabla}^{\rho}\chi\bar{\nabla}_{\rho}\lambda-\bar{\square}\lambda\right)-3V_{\lambda}=0.\label{eq:lambda-FE}
\end{gather}
The shorthand notations $V_{\chi}=\frac{\partial V}{\partial\chi}$
and $V_{\lambda}=\frac{\partial V}{\partial\lambda}$ were used.

\section{Inflation on the FLRW background \label{sec:Inflation}}

Generically, inflation may be regarded as an early period of near-exponential
accelerated expansion taking place at some point roughly in between
$10\text{ MeV}$ and $10^{16}\text{ GeV}$. The motivations for this
early vertiginous expansion range from the need to explain the observed
flat universe, to the attempt to justify the high degree of homogeneity
and isotropy displayed by the CMB, and, more importantly, to predict
the causally connected density fluctuations that are correlated to
the large-scale structure in the present-day universe \cite{Guth1981,MukBook,Linde1983}.

Inflationary cosmology addresses basically three points:
\begin{enumerate}
\item Initial conditions leading to the quasi-exponential expansion;
\item The details of the early accelerated regime and its connections with
observations;
\item The ending of the accelerated expansion and reheating.
\end{enumerate}
The first point is addressed in two ways. Approach number one is more
thorough; it admits a broad range of possible initial conditions in
a non-homogeneous and anisotropic spacetime. The second approach to
point number 1 is a simplified approach assuming generic initial conditions
while the spacetime is restricted to being described by FLRW line
element at the background level \cite{Brand2016}. Notice that the
flatness problem and the problem of generating the primordial fluctuations
can be treated via both the above approaches; however, the problem
of explaining homogeneity and isotropy can only be addressed by the
first, more complete approach. Regardless the approach, a robust inflationary
model should be able to produce an accelerated expansion from fairly
general initial conditions.

Point number 2 is the most relevant one since it directly connects
inflation to observations. In fact, initial fluctuations are generated
during the inflationary dynamics; these density perturbations are
the very seeds of the universe's large-scale structure. The necessary
condition for achieving an inflationary regime accommodating causally
connected perturbation is an accelerated expansion:\footnote{The Hubble function is defined as usual: $H=\dot{a}/a$, where an
overdot denotes differentiation with reespect to the cosmic time $t$.}
\[
\text{Inflation}\Longleftrightarrow\ddot{a}>0\Longleftrightarrow\frac{d}{dt}\left(aH\right)^{-1}<0.
\]
The scale $\left(aH\right)^{-1}$ is known as Hubble horizon or Hubble
radius; it delimits the region wherein two points are momentarily
causally connected. The Hubble radius decreases during inflation allowing
quantum fluctuations to exit the the horizon. These initially correlated
perturbations are then frozen, and later produce the necessary conditions
for structure formation \cite{Baumann2021} (after horizon crossing
at the end of the accelerated period).

Point number 3 addresses the end of the inflationary regime. The particles
the eventually populated the primeval universe were diluted to such
a degree during the almost-exponential expansion that any hint of
a thermalized universe disappears after inflation. Hence, a viable
inflationary model should be able to repopulate the universe after
its ending, then producing a hot Big Band phase dominated by radiation
(ultra-relativistic particles). The period bridging inflation to a
radiation-dominated era is called reheating \cite{Aminetal2014,Lozanov2019}.

The three points above can be (partially) studied in a Friedmann-Lemaître-Robertson-Walker
(FLRW) background. The homogeneous and isotropic FLRW flat spacetime
is described by the line element
\begin{equation}
ds^{2}=-dt^{2}+a^{2}\left(t\right)\left(dx^{2}+dy^{2}+dz^{2}\right),\label{eq:FRLW}
\end{equation}
where $a\left(t\right)$ is the scale factor; natural units are assumed:
$c=1$. Specifying the field equations (\ref{eq:Einstein-FE}), (\ref{eq:chi-FE}),
and (\ref{eq:lambda-FE}) on the spacetime (\ref{eq:FRLW}), leads
to:
\begin{gather}
h^{2}=\frac{1}{2}\left(\frac{1}{2}\chi_{t}^{2}-\frac{\beta_{0}}{6}e^{-\chi}\lambda_{t}^{2}+\bar{V}\left(\chi,\lambda\right)\right),\label{eq:h-Friedmann-Eq}\\
h_{t}=-\frac{3}{4}\chi_{t}^{2}+\frac{1}{4}\beta_{0}e^{-\chi}\lambda_{t}^{2},\label{eq:h-dot}
\end{gather}
and
\begin{gather}
\chi_{tt}+3h\chi_{t}-\frac{\beta_{0}}{6}e^{-\chi}\lambda_{t}{}^{2}+\bar{V}_{\chi}=0,\label{eq:chi-ddot}\\
\beta_{0}e^{-\chi}\left[\lambda_{tt}-\left(\chi_{t}-3h\right)\lambda_{t}\right]-3\bar{V}_{\lambda}=0.\label{eq:lambda-ddot}
\end{gather}
For convenience, the above equations were written in terms of the
dimensionless Hubble function $h$ and the dimensionless potential
$\bar{V}$:

\begin{equation}
h\equiv\frac{1}{\sqrt{\kappa_{0}}}\frac{\dot{a}}{a}\quad\text{and}\quad\bar{V}\left(\chi,\lambda\right)\equiv\frac{1}{\kappa_{0}}V\left(\chi,\lambda\right).\label{eq:h-and-V-bar}
\end{equation}
Moreover, use is made of the dimensionless time derivative
\begin{equation}
A_{t}\equiv\frac{1}{\sqrt{\kappa_{0}}}\dot{A}.\label{eq:A-dot}
\end{equation}

The following three subsection deal with particular solutions to Eqs.
(\ref{eq:h-Friedmann-Eq}), (\ref{eq:h-dot}), (\ref{eq:chi-ddot}),
and (\ref{eq:lambda-ddot}).

\subsection{Starobinsky model \label{subsec:Starobinsky}}

\begin{figure}
\begin{center}

\includegraphics[scale=0.5]{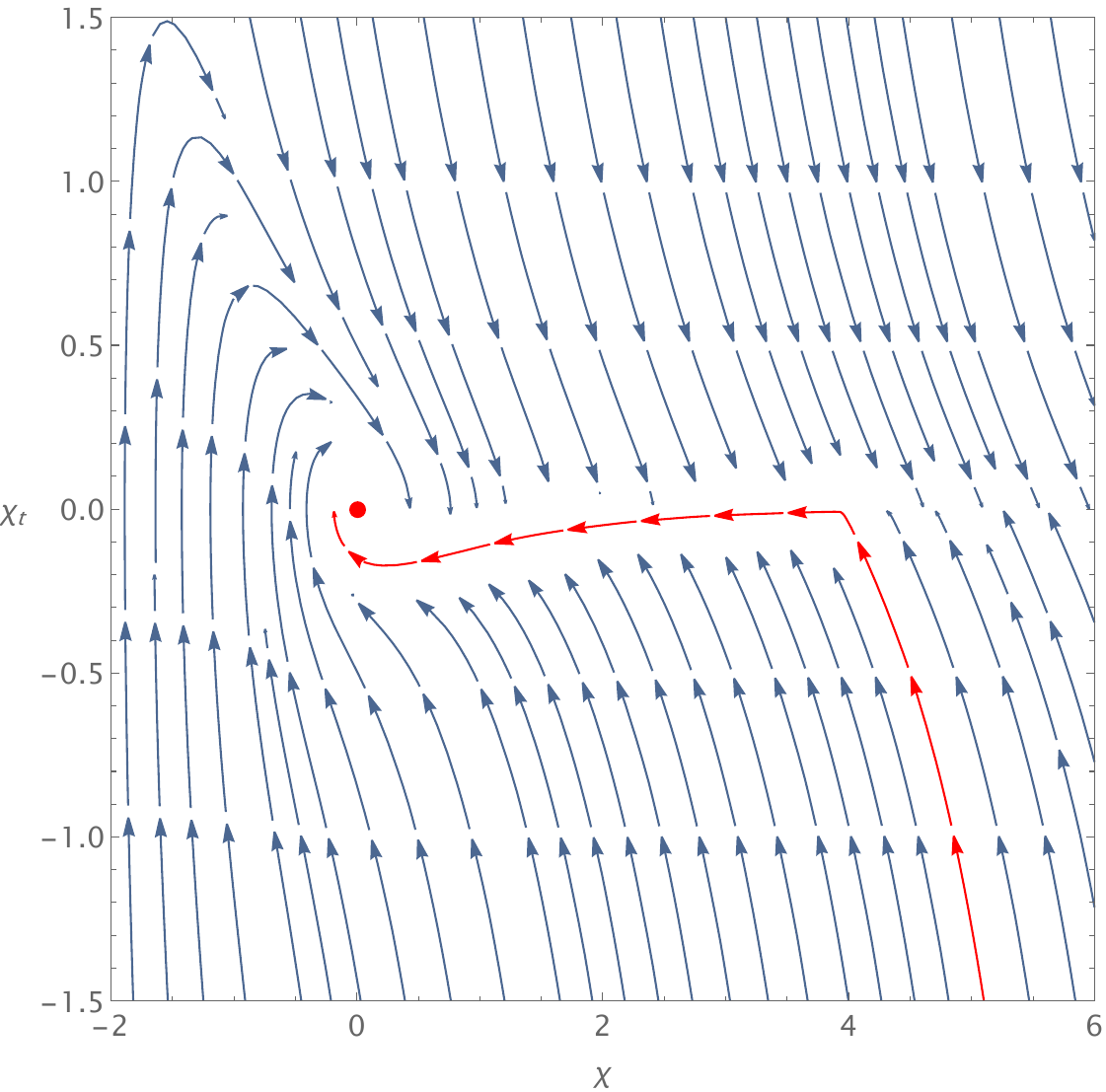}

\end{center}

\caption{Phase space $\left(\chi,\chi_{t}\right)$ for the inflaton field in
Starobinsky model. The red dot corresponds to the accumulation point
at $(0,0)$; the red oriented line highlights a possible trajectory
in the phase space.}
\label{fig:St}

\end{figure}

Starobinsky inflation adds the first-order correction to EH action
via the term proportional $R^{2}$. In this case, $S$ is simplified
by taking $\alpha_{0}=\beta_{0}=0$. Consequently, the field equation
for $\lambda$---Eq. (\ref{eq:lambda-ddot})---becomes a constraint
equation given by:
\begin{equation}
\bar{V}_{\lambda}=0\Rightarrow\lambda=e^{\chi}-1.\label{eq:lambda-constraint-St}
\end{equation}
Inserting (\ref{eq:lambda-constraint-St}) into Eqs. (\ref{eq:h-Friedmann-Eq}),
(\ref{eq:h-dot}) and (\ref{eq:chi-ddot}), leads to:
\begin{align}
h^{2} & =\frac{1}{2}\left(\frac{1}{2}\chi_{t}^{2}+\bar{V}^{\text{St}}\right),\label{eq:h-Fried-St}\\
h_{t} & =-\frac{3}{4}\chi_{t}^{2},\label{eq:acceleration-eq-St}
\end{align}
and
\begin{equation}
\chi_{tt}+3h\chi_{t}+\bar{V}_{\chi}^{\text{St}}=0.\label{eq:chi-ddot-St}
\end{equation}
The $\chi$-related potential $\bar{V}^{\text{St}}\left(\chi\right)$
for the Starobinsky model (label $^{\text{St}}$) reads
\begin{equation}
\bar{V}^{\text{St}}\left(\chi\right)=\frac{1}{6}\left(1-e^{-\chi}\right)^{2}.\label{eq:V-St}
\end{equation}
Notice that Starobinsky inflation is achieved by the dynamics of the
scalar field $\chi$ alone. This dynamics is obtained from Eqs. (\ref{eq:h-Fried-St})
and (\ref{eq:chi-ddot-St}). By taking $\chi$ as the variable describing
the evolution of the system, one rewrites Eq. (\ref{eq:chi-ddot-St})
in the form:
\begin{equation}
\frac{d\chi_{t}}{d\chi}=\frac{-3\chi_{t}\sqrt{\frac{1}{4}\chi_{t}^{2}+\frac{1}{2}\bar{V}^{\text{St}}}-\bar{V}_{\chi}^{\text{St}}}{\chi_{t}}.\label{eq:phase-space-St}
\end{equation}
The above equation is an autonomous first-order ordinary differential
equation; its structure is studied by means of the direction fields
related to $\left(\chi,\chi_{t}\right)$. Fig. \ref{fig:St} shows
the phase space for system of Eq. (\ref{eq:phase-space-St}).

There are two noticeable features in \ref{fig:St}: an approximately
horizontal attractor line in the vicinity of $\chi_{t}\approx0$ and
an accumulation point at the origin.

The attractor line realizes an (almost-)exponential expansion regime
since $\chi_{t}^{2}\ll\bar{V}^{\text{St}}$ ($\chi_{t}\ll1$ and $\bar{V}^{\text{St}}\sim1/6$)
along this trajectory. In fact, by using these conditions in Eqs.
(\ref{eq:h-Fried-St}) and (\ref{eq:acceleration-eq-St}), we build
a slow-roll parameter $\epsilon$ satisfying
\begin{equation}
\epsilon=-\frac{h_{t}}{h^{2}}=\frac{3\chi_{t}^{2}}{\left(\chi_{t}^{2}+2\bar{V}^{St}\right)}\ll1.\label{eq:slow-roll-St}
\end{equation}
The condition $\epsilon\ll1$ yields the inflationary period because
\begin{equation}
h_{t}\ll h^{2}\Rightarrow h\approx\text{constant}\Rightarrow a\left(t\right)\propto\exp\left(\sqrt{\kappa_{0}}ht\right).\label{eq:a(t)-St}
\end{equation}
Moreover, Fig.\ref{fig:St} makes it transparent that a broad range
of initial conditions ($\chi^{i}>2$ and $\chi_{t}^{i}$ arbitrary)
set the system towards the attractor line. Starobinsky model is therefore
capable of producing an inflationary regime starting from a very general
set of initial conditions.\footnote{Starobinsky inflation is an example of the chaotic inflationary scenario
\cite{Linde1983}.}

The accumulation point at $\left(\chi,\chi_{t}\right)=\left(0,0\right)$
is the point of inflation's end. The dynamics of $\chi$ in the vicinity
of this point is oscillatory. This means that $\chi$ transfers energy
to the matter fields it is coupled with while it oscillates coherently
about the origin. The process just described is known as pre-heating;
it is the initial phase of the reheating, when a large number of matter
particles is produced. Since pre-heating is essentially a non-thermal
process, a subsequent thermalization stage is demanded to lead the
universe to a radiation-dominated era where all kinds of matter particles
are in thermal equilibrium \cite{KofLinSta1997,BaTsuWan2006}.

In spite of being a preliminar analysis, the above study based on
Fig. \ref{fig:St} shows that Starobinsky model successfully addresses
the three basic points of interest listed at the beginning of Section
\ref{sec:Inflation}. In the next two subsection, it will be checked
if that continues to be the case for the models including the $R^{3}$-
and $R\square R$-type corrections to Starobinsky inflation.

\subsection{Starobinsky-plus-$R^{3}$ model \label{subsec:Starobinsky-R3}}

A term of the type $R^{3}$ can be added to Starobinsky action ($\propto R+R^{2}$)
thus generating the Starobinsky-plus-$R^{3}$ model. The inflationary
dynamics accommodated by this modified gravity model respects Eqs.
(\ref{eq:h-Fried-St}), (\ref{eq:acceleration-eq-St}), and (\ref{eq:chi-ddot-St})
provided that $\bar{V}^{\text{St}}$ is generalized to the potential
\cite{PRD2022}
\begin{equation}
\bar{V}^{\alpha_{0}}\left(\chi\right)=\frac{e^{-2\chi}}{72\alpha_{0}^{2}}\left(1-\sqrt{1-4\alpha_{0}\left(1-e^{\chi}\right)}\right)\left(-1+8\alpha_{0}\left(1-e^{\chi}\right)+\sqrt{1-4\alpha_{0}\left(1-e^{\chi}\right)}\right).\label{eq:V-R3}
\end{equation}
The potential is real-valued regardless of the value taken by $\chi$
under the constraint: $0\leq4\alpha_{0}\leq1$.

\begin{figure}
\begin{center}

\includegraphics[scale=0.5]{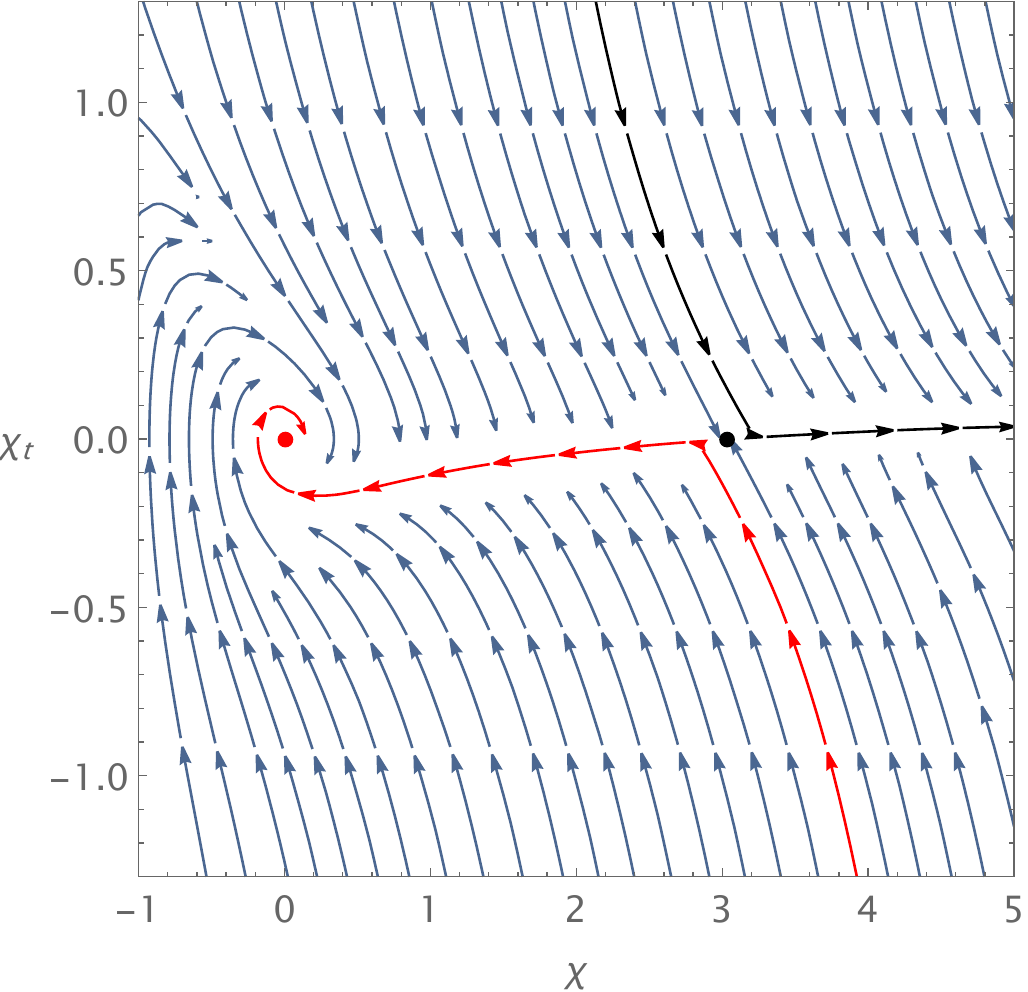}

\end{center}

\caption{Phase-space representation $\left(\chi,\chi_{t}\right)$ for Starobinsky-plus-$R^{3}$
model with parameter $\alpha_{0}=10^{-2}$. The red dot and the black
dot in the plot mark the critical points $\left(0,0\right)$ and $\left(\chi_{c},0\right)$
where $\chi_{c}=3.06$. The red line and the black line show two opposite
trajectories with respect to the critical point $\left(\chi_{c},0\right)$.
Source: Ref. \cite{PRD2022}.}
\label{fig:St-R3}
\end{figure}

The phase-space analysis for the Starobinsky-plus-$R^{3}$ model is
performed along the lines of what was done in Section \ref{subsec:Starobinsky},
by employing Eq. (\ref{eq:phase-space-St}) with the substitution
$\bar{V}^{\text{St}}\rightarrow\bar{V}^{\alpha_{0}}$. This leads
to Fig. \ref{fig:St-R3}.

The main difference between the Figs. \ref{fig:St-R3} and \ref{fig:St}
is the appearance of a new critical point
\begin{equation}
P_{c}=\left(\chi_{c},0\right)=\left(\ln\left(4+\sqrt{3\alpha_{0}^{-1}}\right),0\right).\label{eq:P_c}
\end{equation}
This critical point is a saddle point that splits the phase space
into two distinct regions in regard to the direction field lines.
The sector of Fig. \ref{fig:St-R3} to the left of the vertical red
attractor line yields an inflationary regime ending in the stable
accumulation point $\left(\chi,\chi_{t}\right)=\left(0,0\right)$.
If the inflaton field $\chi$ starts from $\left(\chi^{i},\chi_{t}^{i}\right)$
in this region, inflation occurs in the usual way: the accelerated
expansion subsequently gives off into a decelerated phase with $\chi$
oscillating about the origin (potential minimum). On the other hand,
the trajectories to the right from the vertical black line yield a
inflationary dynamics that never ends. In fact, Ref. \cite{PRD2022}
details how the field $\chi$ grows indefinitely in this sector; its
dynamics leading to the transition of an initial almost-exponential
expansion to the asymptotically accelerated phase of the power-law
type $a\left(t\right)\sim t^{12}$.

The precise location of the point $P_{c}$ is sensitive to the value
of parameter $\alpha_{0}$ accompanying the term $R^{3}$ in the action
(\ref{eq:S}). The smaller the value of $\alpha_{0}$, the greater
the value of $\chi_{c}$. Reassuringly enough, in the (Starobinsky)
limit $\alpha_{0}\rightarrow0$, it is $\chi_{c}\rightarrow\infty$,
and Fig. \ref{fig:St-R3} degenerates into \ref{fig:St}, as it should.
The most notable difference between the Starobinsky-plus-$R^{3}$
setup and the standard Starobinsky inflation is the fact that the
initial conditions leading to a physically meaningful inflation cannot
be chosen arbitrarily.\footnote{By ``physically meaningful inflation'' it is meant a primordial accelerated expansion that ends in a radiation-dominated Hot Big-Bang phase.} In effect, the greater the value of $\alpha_{0}$ the smaller the
set of initial conditions $\left(\chi^{i},\chi_{t}^{i}\right)$ capable
of producing an inflationary regime that evolves to a radiation epoch.
In this sense, a introduction of the $R^{3}$ correction to the Starobinsky
model requires some sort of fine tuning in the initial conditions
of the inflaton field \cite{PRD2022}.

\subsection{Starobinsky-plus-$R\square R$ model\label{subsec:Starobinsky-Rd2R}}

This section deals with the changes to inflation resulting from the
inclusion of a $R\square R$-type correction to Starobinsky model.

This scenario is called the Starobinsky-plus-$R\square R$ model;
its main distinctive feature with respect to the previous cases (Subsections
\ref{subsec:Starobinsky} and \ref{subsec:Starobinsky-R3}) is the
presence of two scalar fields $\chi$ and $\lambda$ that are both
responsible for the background dynamics, i.e. wherein a multi-field
inflation will be realized. The related phase-space analysis is performed
rewriting the second-order equations (\ref{eq:chi-ddot}) and (\ref{eq:lambda-ddot})
as a system of four first-order equations. Accordingly, by defining
\begin{equation}
\chi_{t}=\psi\quad\text{and}\quad\lambda_{t}=\phi\label{eq:psi_and_phi}
\end{equation}
it results:
\begin{align}
\chi_{t} & =\psi,\label{eq:autonomous-psi}\\
\psi_{t} & =-3h\psi+\frac{\beta_{0}}{6}e^{-\chi}\phi^{2}-\bar{V}_{\chi}^{\beta_{0}},\label{eq:autonomous-psi-dot}\\
\lambda_{t} & =\phi,\label{eq:autonomous-phi}\\
\beta_{0}\phi_{t} & =\beta_{0}\left(\psi-3h\right)\phi+3e^{\chi}\bar{V}_{\lambda}^{\beta_{0}},\label{eq:autonomous-phi-dot}
\end{align}
with
\begin{equation}
h=\sqrt{\frac{1}{2}\left(\frac{1}{2}\psi^{2}-\frac{\beta_{0}}{6}e^{-\chi}\phi^{2}+\bar{V}^{\beta_{0}}\right)}\label{eq:h-HOSt}
\end{equation}
and
\begin{equation}
\bar{V}^{\beta_{0}}\left(\chi,\lambda\right)=\lim_{\alpha_{0}\rightarrow0}\bar{V}\left(\chi,\lambda\right)=\frac{1}{3}e^{-2\chi}\lambda\left(e^{\chi}-1-\frac{1}{2}\lambda\right).\label{eq:V-HOSt}
\end{equation}
Notice that: (i) the phase space is four dimensional in the higher-order
Starobinsky model---it is built with $\chi$, $\chi_{t}$, $\lambda$,
and $\lambda_{t}$; (ii) the dimensionless Hubble function $h$ depends
explicitly on the parameter $\beta_{0}$---the coupling of the $R\square R$-term
in the action (\ref{eq:S}); and, predictably (iii) the dimensionless
potential $\bar{V}^{\beta_{0}}$ is a double-field quantity.

The autonomous system formed by Eqs. (\ref{eq:autonomous-psi}), (\ref{eq:autonomous-psi-dot}),
(\ref{eq:autonomous-phi}), and (\ref{eq:autonomous-phi-dot}) admits
a single critical point at the origin:
\begin{equation}
P_{0}=\left(\chi_{0},\lambda_{0},\psi_{0},\phi_{0}\right)=\left(0,0,0,0\right).\label{eq:P_0}
\end{equation}
Eqs. (\ref{eq:autonomous-psi})---(\ref{eq:autonomous-phi-dot})
can be linearized about point $P_{0}$. Thereby, it follows that Lyapunov
exponents $r_{0}$ related to the stability of the critical point
satisfy the algebraic equation
\begin{equation}
\beta_{0}r_{0}^{4}+r_{0}^{2}+\frac{1}{3}=0.\label{eq:Lyapunov-eq}
\end{equation}
The solution of Eq. (\ref{eq:Lyapunov-eq}),
\begin{equation}
r_{0}=\pm\sqrt{\frac{-1\pm\sqrt{1-\frac{4\beta_{0}}{3}}}{2\beta_{0}}},\label{eq:r_0}
\end{equation}
and the analysis of the direction fields in the phase-space, lead
to the conclusion that $P_{0}$ is a stable fixed point only within
the interval
\begin{equation}
0\leq\beta_{0}\leq\frac{3}{4}.\label{eq:beta-constraint}
\end{equation}
In fact, any value of $\beta_{0}$ outside the above interval leads
to $\text{Re}\left[r_{0}\right]>0$ for at least one of the four possible
$r_{0}$ in Eq. (\ref{eq:r_0}). To put it another way, the equilibrium
point $P_{0}$ is unstable whenever condition (\ref{eq:beta-constraint})
is violated. Moreover, it is worth mentioning that the stability of
$P_{0}$ is a necessary condition for a graceful exit from inflation
into a radiation-dominated universe. This result was first published
in Ref. \cite{JCAP 2019} and later reanalyzed by \cite{EPJC2023}.

Details about the dynamics of the double-field higher-order Starobinsky
inflation are obtained from the numerical analysis of the four-dimensional
phase space $\left(\chi,\chi_{t},\lambda,\lambda_{t}\right)$. For
this end, Eqs. (\ref{eq:chi-ddot}) and (\ref{eq:lambda-ddot}) are
cast into the form:
\begin{gather}
\frac{d\chi_{t}}{d\chi}=\frac{-3h\chi_{t}+\frac{\beta_{0}}{6}e^{-\chi}\lambda_{t}^{2}-\bar{V}_{\chi}^{\beta_{0}}}{\chi_{t}},\label{eq:phase-space-chi}\\
\frac{d\lambda_{t}}{d\lambda}=\left(\chi_{t}-3h\right)+\frac{3e^{\chi}}{\beta_{0}\lambda_{t}}\bar{V}_{\lambda}^{\beta_{0}},\label{eq:phase-space-lambda}
\end{gather}
with $h$ given by Eq. (\ref{eq:h-HOSt}).

\begin{figure}
\begin{center}

\includegraphics[scale=0.5]{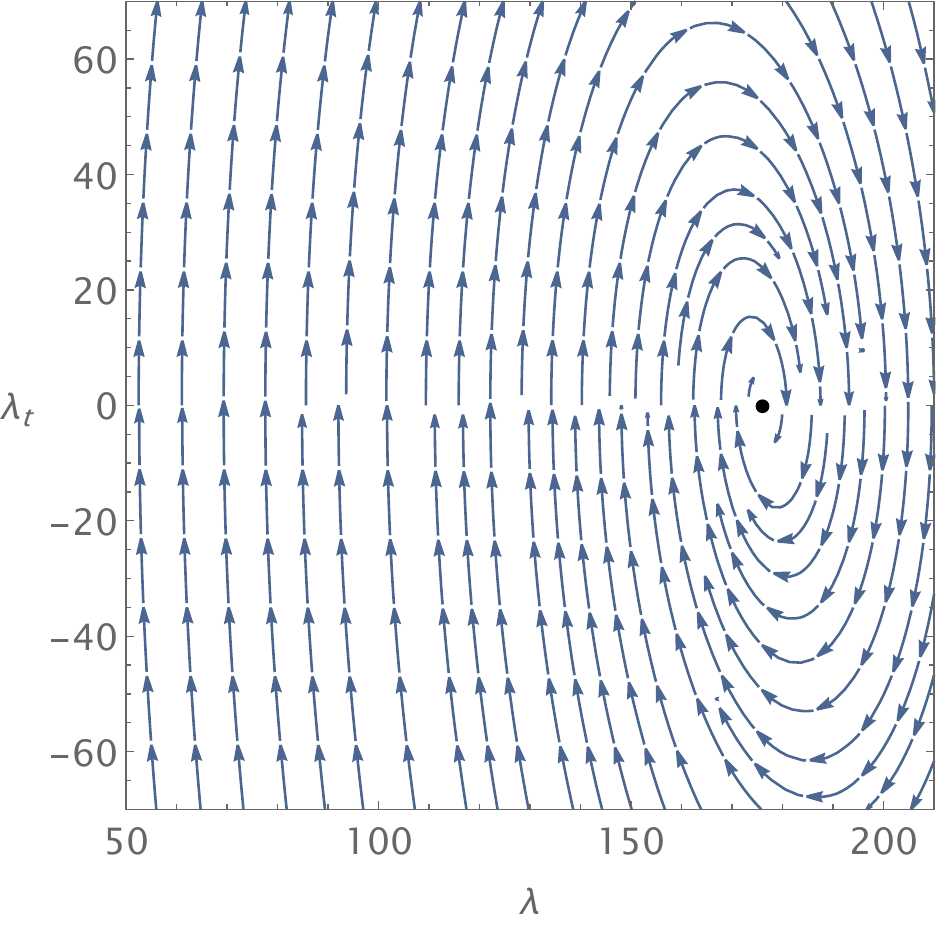}$\qquad$\includegraphics[scale=0.5]{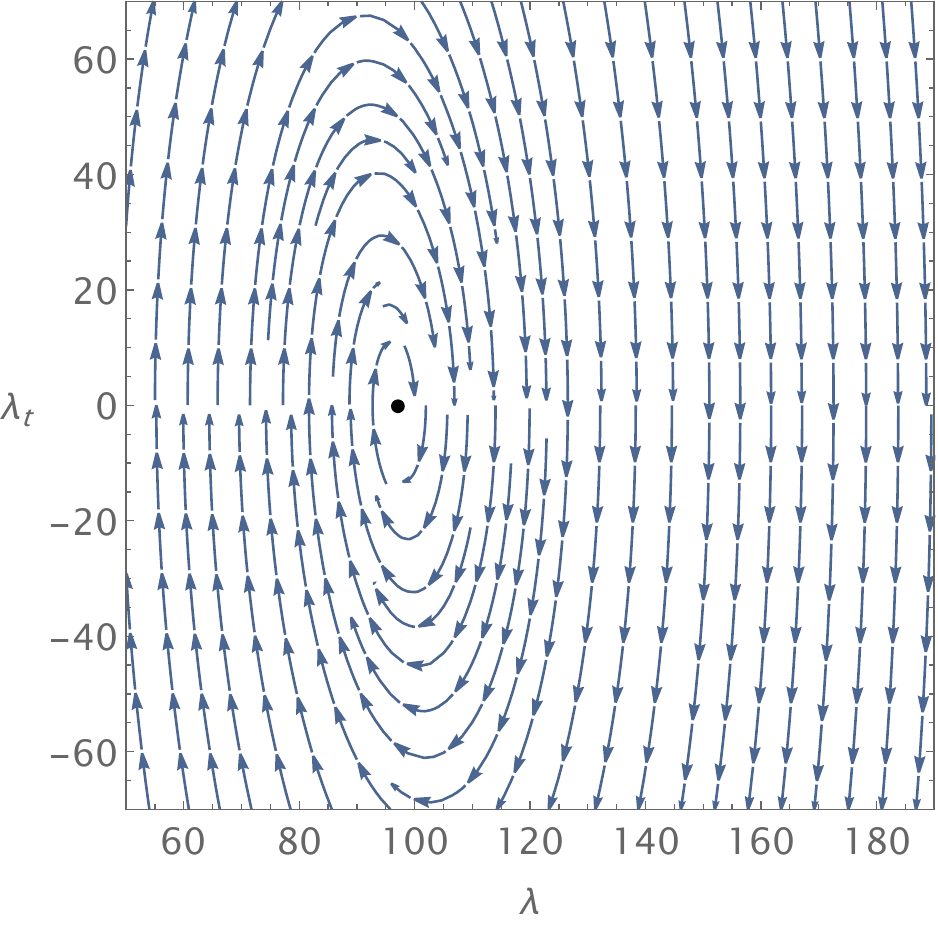}

\end{center}

\caption{Two-dimensional $\left(\lambda,\lambda_{t}\right)$-slices of the
four-dimensional $\left(\chi,\chi_{t},\lambda,\lambda_{t}\right)$-phase
space corresponding to $\beta_{0}=10^{-3}$ and $\chi_{t}=0$ with
$\chi=5.18$ (left panel) and $\chi=4.58$ (right panel). The black
dots mark the position of the accumulation points: $\left(\lambda,\lambda_{t}\right)\simeq\left(177,0\right)$
in the left panel; $\left(\lambda,\lambda_{t}\right)\simeq\left(98,0\right)$
in the right panel.}
\label{fig:HOSt}
\end{figure}

By using Eqs. (\ref{eq:phase-space-chi}) and (\ref{eq:phase-space-lambda}),
two-dimensional slices of the phase space can be performed, e.g. plots
of $\left(\chi,\chi_{t}\right)$ and $\left(\lambda,\lambda_{t}\right)$
are built for fixed values of $\beta_{0}$ and of the remaining dynamical
variables. Specifically, the direction fields in the $\left(\chi,\chi_{t}\right)$
slice are obtained by choosing adequate values for $\beta_{0}$, $\lambda$,
and $\lambda_{t}$; Fig. \ref{fig:St} is representative of the $\left(\chi,\chi_{t}\right)$
plane thus constructed: it is verified the existence of an attractor
line close to $\chi_{t}\simeq0$ in the Starobinsky-plus-$R\square R$
model. The attractor line in the $\left(\chi,\chi_{t}\right)$-plane
is very robust in the sense that it exists for arbitrary values of
$\beta_{0}$, $\lambda$, and $\lambda_{t}$ that are consistent with
a physical inflation---i.e. $\beta_{0}$ within the interval in (\ref{eq:beta-constraint})
and a real-valued $h$ given by Eq. (\ref{eq:h-HOSt}). On the other
hand, the direction fields for the two-dimensional slice $\left(\lambda,\lambda_{t}\right)$
are built by fixing the values assumed by $\beta_{0}$, $\chi$, and
$\chi_{t}$; Fig. \ref{fig:HOSt} illustrates two such examples of
$\left(\lambda,\lambda_{t}\right)$-plane slices.

A joint analysis of Figs. \ref{fig:St} and \ref{fig:HOSt} indicates
that the field $\chi$ approaches the attractor line $\chi_{t}\simeq0$
simultaneously as the field $\lambda$ tends to the accumulation point
where $\lambda_{t}\rightarrow0$ and $\lambda\simeq e^{\chi}$. This
attractor trajectory $\left(\chi,\lambda,\chi_{t}\text{,}\lambda_{t}\right)\simeq\left(\chi,e^{\chi},0,0\right)$
in the four-dimensional phase space corresponds to the configuration
realizing the inflationary regime. This fact is verified from the
first-order approximation slow-roll parameter
\begin{equation}
\epsilon\simeq\frac{4e^{-2\chi}}{\left(3-\beta_{0}e^{\chi}\right)}.\label{eq:slow-roll-HOSt}
\end{equation}
Internal consistency with the first-order approximations requires
$\beta_{0}e^{\chi}$ smaller than (but not to close to) $3$. Accordingly,
Eq. (\ref{eq:slow-roll-HOSt}) shows that the inflationary regime
$\left(\epsilon\ll1\right)$ takes place whenever $\chi\gtrsim2$.
Further details on the show-roll regime are available in Ref. \cite{EPJC2023}.

We summarize the analysis of this subsection by stating the three
basic conditions that must be satisfied for achieving a physical inflationary
regime within the Starobinsky-plus-$R\square R$ model: (1) Parameter
$\beta_{0}$ should pertain to the interval of values specified in
(\ref{eq:beta-constraint}); (2) The initial condition for the field
$\chi$ must comply with $\chi^{i}\gtrsim2$; and (3) The dimensionless
Hubble function should be well defined, i.e. $h\left(t\right)\in\mathbb{R}$,
for all trajectories taken by the fields $\chi$ and $\lambda$.

\section{Final remarks \label{sec:Conclusion}}

This article recalls some of the motivations to consider extensions
to general relativity for describing the gravitational interaction.
A particular branch of modified gravity proposals is chosen as the
focus, namely that of higher-order gravity. The latter admits a four-dimensional
Riemaniann manifold endowed with a rank-2 metric tensor $g_{\mu\nu}$
which field equations include derivatives of order higher than two
(Section \ref{sec:Introduction}). For this reason, Einstein-Hilbert
action (wherein the Lagrangian density is $\mathcal{L}^{\text{EH}}\propto R\sim\partial^{2}g$)
is generalized into Starobinsky model ($\mathcal{L}^{\text{St}}\propto R+R^{2}$),
and further into the higher-order Starobinsky model ($\mathcal{L}^{\text{HOSt}}\propto R+R^{2}+\alpha_{0}R^{3}+\beta_{0}R\square R$)---cf.
Section \ref{sec:Higher-order-model}. The main scope of the paper
was to specify the field equations for $g_{\mu\nu}$ and the extra
scalar degree(s) of freedom $\chi$ (and $\lambda$) for the homogeneous
and isotropic FLRW background of non-perturbative cosmology before
studying the early-universe inflationary regime allowed within those
modified gravity models (Section \ref{sec:Inflation}).

Three specific examples were scrutinized in Subsections \ref{subsec:Starobinsky}
through \ref{subsec:Starobinsky-Rd2R}. Starobinsky model was taken
as the paradigm of successful realization of inflation. Its dynamics
was studied carefully in the phase space because of its transparency
and for setting the stage for the more complicated models that followed.
Starobinsky's inflaton field dynamics follows an attractor line towards
an accumulation point for arbitrary general initial conditions (Fig.
\ref{fig:St}). It engenders a quasi-exponential expansion that exits
gracefully to a radiation-dominated universe.

The same possibility---that of an inflation ending in a Hot Big-Bang
universe---is realized within the Starobinsky-plus-$R^{3}$ model
(Subsection \ref{subsec:Starobinsky-R3}), albeit for a more restrict
set of initial conditions. In fact, besides the accumulation point
at the origin of the phase space $\left(\chi,\chi_{t}\right)$ for
the single inflaton field of this model, there is an unstable equilibrium
point depending on $\chi_{c}=\ln\left(4+\sqrt{3\alpha_{0}^{-1}}\right)$;
the associated trajectories split the phase space into two region,
one of those leading to eternal inflation (see Fig. \ref{fig:St-R3}).
In order to tone down this possibility, the parameter $\alpha_{0}$
could be constraint to assume small values.

Parameter $\beta_{0}$ typical of the Starobinsky-plus-$R\square R$
model is also constrained based on similar arguments. However, this
case is more evolved partially due to the fact that there are two
scalar degrees of freedom ($\chi$ and $\lambda$) playing the role
of the inflaton. The related double-field inflation is achieved by
requiring $\beta_{0}$ to take on values within the interval $0\leq\beta_{0}\leq3/4$.
This requirement is based both on the phase-space analysis and on
the demand for stability of the critical point at the origin of the
four-dimensional space $\left(\chi,\lambda,\chi_{t},\lambda_{t}\right)$.
Subsection \ref{subsec:Starobinsky-Rd2R} contains the details of
how to slice the phase space into two-dimensional sectors (such as
those in Fig. \ref{fig:HOSt}) leading to these conclusions and to
the additional requirement that it should be $\chi^{i}\gtrsim2$ for
an initial condition leading to a physical inflationary regime.

The constraints on the parameters $\alpha_{0}$ and $\beta_{0}$ deduced
here stem from a simplistic reasoning based on the backgroung evolution
of the field equations. These constraints can be refined by the perturbative
treatment of the modified gravity models. This technically sofisticated
task is undertaken elsewhere---see e.g. \cite{EPJC2023,PRD2022,JCAP 2019}.
In fact, the CMB data offers a contour region in a plot of the tensor-to-scalar
ratio $r$ in as a function of the scalar tilt $n_{s}$ \cite{BICEP3}.
Starobinsky model is highly favored because its prediction for $r=r\left(n_{s}\right)$
for a number of e-folds in the interval $50\lesssim N\lesssim60$
respects $r\lesssim0.01$. Starobinsky-plus-$R^{3}$ model \cite{PRD2022}
and Starobinsky-plus-$R\square R$ model \cite{EPJC2023,JCAP 2019}
are also consistent with CMB observations; additionally, they allow
for a larger variability of $n_{s}$ values thus accommodating a greater
flexibility for data constraining. This might be consistent with the
possibility of a non-null running of $n_{s}$ in the power spectrum
parameterization \cite{Planck2020}.

\section*{Acknowledgements}

RRC and LGM are grateful to Ruben Aldrovandi for his supervision,
his friendship, the inspiration, and the time they have shared at
IFT-UNESP. The authors thank CNPq-Brazil for partial financial support---Grants:
309984/2020-3 (RRC) and 307901/2022-0 (LGM).

\end{document}